\documentclass[aps,pra,twocolumn,showpacs,groupedaddress,amsmath,amssymb]{revtex4}

\usepackage{graphicx}
\usepackage{color}
\usepackage{comment}
\usepackage{bm}
\usepackage{latexsym}

\begin{document}

\title{Design algorithms of driving-induced nonreciprocal components}
\author{Huanan Li, and Tsampikos Kottos} 
\affiliation{Wave Transport in Complex Systems Lab, Department of Physics, Wesleyan University, Middletown, CT-06459, USA}
\date{\today}

\begin{abstract}
We utilize an effective Hamiltonian formalism, within the Floquet scattering framework, to design a class of driving-induced non-reciprocal 
components with {\it minimal} complexity. In the high driving-frequency limit, where our scheme is formally applicable, these designs 
demonstrate a leading order non-reciprocal performance which is inverse proportional to the driving frequency. Surprisingly, the optimal 
non-reciprocal behavior persists also in the slow driving regime. Our approach highlights the importance of physical loops in the design 
of these driven non-reciprocal components.
\end{abstract}

\maketitle
\section{Introduction}\label{intro}

The quest for schemes that lead to the realization of novel non-reciprocal components has been, for many years 
now, a subject of intense activity \cite{LRWZHL12,Alu0,mezler,Alu1,Alu2, Alu3,YF11,SYF15}. On the fundamental level these schemes 
must invoke mechanisms that violate time-reversal symmetry -- the latter being in the core of wave transport reciprocity theorems 
\cite{LL60,C45,H84,ST07}. On the technological level, the employed designs must satisfy a set of requirements; the non-reciprocal 
components have to be bounded by the small footprint of the devices, they have to be easily fabricated, have small cost, small energy 
consumption etc. Needless to say that any achievement along these lines can influence the progress of wave transport management 
in areas as diverse as electromagnetism, acoustics, thermal, matter and quantum waves. At the same time, the development of such 
new class of circulators, isolators and other non-reciprocal components will have dramatic implications in the next generation of 
communication, protection, imaging, and quantum information schemes.

In the electromagnetic framework, non-reciprocal transport has been mainly achieved via magneto-optical materials \cite{ZK97,
DBZHWGHP05}. These exotic materials are normally very lossy when deposited at thin films. Moreover, the weak nature of the 
magneto-optical effects makes them incompatible with on-chip integration. Similar problems appear in acoustics, where 
magneto-acoustic effects are even weaker than their optical counterparts \cite{L05}. An alternative approach to directional wave transport 
utilizes nonlinear spatially asymmetric structures \cite{Alu0,mezler,BFBRCEK13,NBRMCK14,LC11,Yang14,LYC09,LGTZ10,BTD11}. 
The nonlinear effects can be different for the forward and backward propagating waves, thus resulting in intensity-dependent propagation 
asymmetry. The same principle applies also for phononic heat transport and can lead to thermal diodes and rectifiers \cite{mezler,
LRWZHL12,WL08,LWC04}. Despite this success, nonlinear mechanisms impose limitations on the operational 
amplitude of the device - an undesirable feature from the engineering perspective. At the same time, in the case of electromagnetic 
and acoustic waves, they often introduce inherent signal distortions (generation of higher harmonics) \cite{mezler}. 

\begin{figure}
\includegraphics[width=1\linewidth, angle=0]{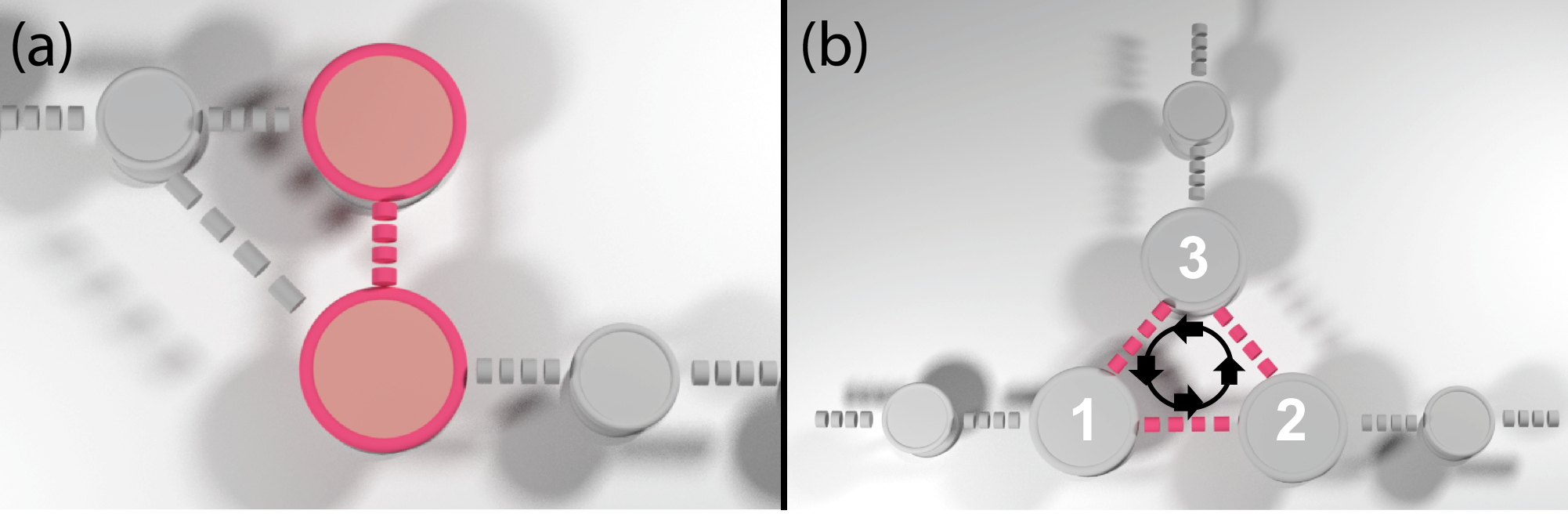}
\caption{Non-reciprocal components with minimal design complexity. Red highlights indicate the components which are potentially 
modulated in time. (a) An isolator based on two-mode modulated target.  In the specific case, they can be the resonant frequency 
of each mode and the coupling between them. (b) A circulator based on a three-mode target where the coupling between the modes
is periodically modulated. The bold circle indicates the direction of circulation. 
}
\label{fig1}
\end{figure}

Parallel approaches that aim to realize non-reciprocal components have capitalized on active schemes. These designs utilize 
spatio-temporal modulations of the impedance profile of the propagating medium and provide a promising alternative for the 
realization of compact, reconfigurable non-reciprocal components \cite{Alu0,Alu1,Alu2,FSSHA14,YF11,LK15,SLFK16}. In fact, 
when paired with the emerging field of {\it Floquet engineering} \cite{Goldman2014,Eckardt2015,IK15}, they might provide a powerful 
approach that can produce frequency and bandwidth-tailored non-reciprocal transport \cite{Li2018,LiBT2018}.

In this paper we will utilize the toolbox of Floquet engineering, in order to design a class of driving-induced non-reciprocal components 
with {\it minimal} complexity. Our design scheme is demonstrated using the universal framework provided by coupled mode theory. 
Although it is formally applicable at the high-frequency modulation limit, it also provide guidance for the design of non-reciprocal 
components in the low-frequency modulation regime. The proposed methodology highlights the importance of physical loops 
in these designs. Moreover, it allow us to derive analytical expressions for the left/right transmittance asymmetry $\Delta=
T_{L\rightarrow R}-T_{R\rightarrow L}$ which, in the high-frequency $\omega$ modulation regime, is $\Delta\sim {\cal O}(1/\omega)$. 
We find that the theoretical expressions that describe the transmittance asymmetry $\Delta$ of this class of isolators and circulators 
matches nicely with the numerical results and in many occasions can reach the maximum value of $100\%$ non-reciprocal behavior.

The structure of the paper is as follows. In Sec. \ref{scattering}, we present the Floquet engineering scattering formalism in the limit 
of high modulation frequency. The method utilizes the notion of effective Floquet Hamiltonian. In the next Sec. \ref{high}, we implement 
this formalism for the design of reconfigurable non-reciprocal components with leading order performance which is inverse proportional
to the driving frequency. In Subsec. \ref{isolators}, we discuss the applicability of our scheme for the case of an isolator and demonstrate 
the validity of our approach via a specific model. The applicability of the method in the case of a circulator is shown in Subsec. 
\ref{circulators}. In Sec. \ref{low}, we demonstrate the persistence of optimal performance of our designs in the low driving frequency 
regimes via numerical examples. Finally, our conclusions are given at the last Sec. \ref{conclusions}.

\section{Floquet scattering in the high modulation frequency limit} \label{scattering}

\begin{figure}
\includegraphics[width=1\linewidth, angle=0]{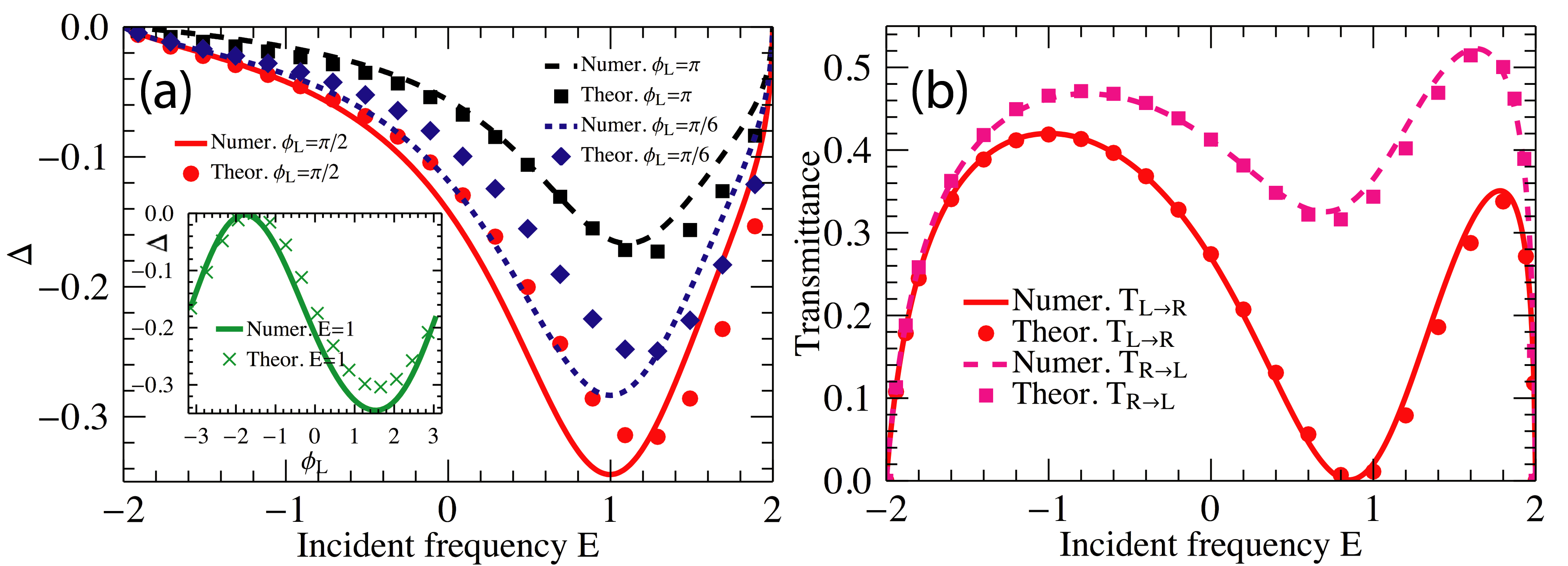}
\caption{Non-reciprocal characteristics of a two-mode driven isolator with a physical loop (see Fig. \ref{fig1}a). The driving protocol is 
given by Eq. (\ref{Eg_isolator}). (a) The numerical (lines) values (using the Floquet scattering matrix approach, see Ref.~\cite{Li2018}), 
for the transmittance asymmetry $\Delta$ versus the incident frequency $E$, are compared with the theoretical (symbols) results of 
Eq.~(\ref{isolator_delta}). Three different relative driving phases for the left resonator $\phi_{L}=\pi,$ $\pi/2$ and $\pi/6$ are studied. The 
best isolation performance is observed when $\phi_{L}=\pi/2$ (see also the inset). Inset: the transmittance asymmetry $\Delta$ versus the 
phase $\phi_{L}$ when the incident frequency is fixed to be $E=1$.  (b) Left/right transmittances ($T_{L\rightarrow R}/T_{R\rightarrow L}$) 
versus the incident frequency $E$ when the optimized phase $\phi_{L}=\pi/2$ is used. The theoretical result (symbols) are given by 
Eq.~(\ref{Smat}). In both (a) and (b) we have used the following set of parameters associated with the driving protocol Eq. (\ref{Eg_isolator}): $\varepsilon_{L}^{0}=\varepsilon_{R}^{0}=0$, $f_{L}=f_{R}=1$, $\phi_{R}=-\pi/2$, $h_{0}=-2h=-1$, $\omega=6$ and finally $\gamma=0.3$ 
for the uniform loss. The bare coupling matrix that describes the coupling between the target and the leads has matrix elements $c_{L}=
c_{R}=\tilde{c}_{L}=-1$,
\label{fig2} }
\end{figure}

We consider periodically time-modulated systems consisting of $N_s$ 
coupled modes. In the context of coupled mode theory such a (hermitian) system can be described by a time-dependent $N_s$-dimensional 
Hamiltonian $\hat{H}_{0}\left(t\right)=\hat{H}_{0}^{\dagger}\left(t\right)=\hat{H}_{0}\left(t+T\right)$. It turns out that the evolution of such
systems can be expressed in terms of a $N_s$-dimensional time-independent effective (Floquet) Hamiltonian $\hat{H}_{F}$ and a 
micromotion operator $\hat{U}_{F}\left(t\right)=\hat{U}_{F}\left(t+T\right)$ \cite{Goldman2014}. The former describes the long time 
(stroboscopic) dynamics while the latter accounts for the evolution within one period of the driving. This formalism allows us to invoke a 
systematic high-frequency expansion in $\omega=2\pi/T$ for both the effective Hamiltonian $\hat{H}_{F}$ and the micromotion operator 
$\hat{U}_{F} \left(t\right)$ \cite{Eckardt2015}. A benefit of this representation is the fact that one can eliminate the artificial dependence 
of the effective Hamiltonian on the driving phase (i.e. initial time of periodic driving) which can be elusive as far as symmetry preservation 
is concerned \cite{Eckardt2015}.

At the same time, the notion of $\hat{H}_{F}$ and $\hat{U}_{F}$ has been proven useful in the analysis of Floquet scattering settings 
\cite{LiBT2018}. The latter constitutes the natural framework where the scattering properties of driven targets can be inferred from the 
dynamics of their isolated (i.e. in the absence of coupling with leads) counterparts \cite{Li2018}. The proposed approach \cite{LiBT2018} 
allows us to treat the driven target as a static one -- which is described by the effective Hamiltonian $\hat{H}_{F}$ -- coupled to leads. In 
this framework the coupling constants that describe the coupling of the target with the leads are also time-independent - albeit they are 
different from their bare values due to a renormalization procedure which involves information encoded in the micromotion. 

To be specific, we consider a time-periodic driven target which is coupled with $M$ identical leads. The coupling strengths between the 
target and the leads are given by the bare coupling matrix $\hat{c}$. The transport characteristics of this system are given by the Floquet 
scattering matrix $\mathcal{S}$, 
which couples the outgoing propagating channels with the incoming ones. Below we will assume for simplicity that in the high-driving 
frequency domain there exists only one propagating channel in each lead. The case of multimode leads, although more cumbersome,
can be also worked out along the same lines. The flux-normalized scattering matrix $S$, up to order ${\cal O}\left(1/\omega^{2}\right)$, 
is \cite{LiBT2018}
\begin{align}
S & \approx-I_{M}+\imath v_{g}\hat{c}_{u}G\hat{c}_{u}^{\dagger},\, 
G=\frac{1}{E-\hat{H}_{F}+\varLambda_{u}+\frac{\imath}{2}v_{g}\hat{c}_{u}^{\dagger}\hat{c}_{u}}
\label{Smat}
\end{align}
where $I_{M}$ is the $M\times M$ identity matrix, $v_{g}=\partial E/\partial k$ is the group velocity of the propagating channel, and 
$\hat{c}_{u}=\hat{c}\hat{u}_{0}$ is the renormalized bare coupling due to the micromotion $\hat{u}_{n}\equiv\frac{1}{T}\int_{0}^{T}
dt\hat{U}_{F}\left(t\right)e^{\imath n\omega t}$. The latter can be approximated as
\begin{equation}
\hat{u}_{n}\approx\left\{\begin{array}{cc}
I_{N_s}-\frac{1}{2\omega^{2}}\sum_{n\neq0}\frac{1}{n^{2}}\hat{H}^{\left(-n\right)}\hat{H}^{\left(n\right)};& \quad n=0 \\
\frac{1}{n\omega}\hat{H}^{\left(n\right)};&\quad n\neq 0 \label{u0}
\end{array}\right.
\end{equation}
where $\hat{H}^{\left(n\right)}\equiv\frac{1}{T}\int_{0}^{T}dt\hat{H}_{0}\left(t\right)e^{\imath n\omega t}$. The term $\varLambda_{u}
=\lambda\hat{c}_{u}^{\dagger}\hat{c}_{u}+\sum_{n}\hat{u}_{n}^{\dagger}\imath\hat{\Gamma}\hat{u}_{n}$ in the denominator of Eq. 
(\ref{Smat}) represents the fact that the target is not isolated. The origin of these two terms is different: the first one describes the 
channel-coupling induced renormalization to the close-system Hamiltonian while the second one is optional and describes potential 
material gain/loss of the target ($-\imath\hat{\Gamma}$ is a non-Hermitian diagonal matrix). In fact, in an alternative formulation of 
the problem, we could absorb the gain/loss properties of the isolated system in a composite coupled mode Hamiltonian $\hat{H}_0
(t)\rightarrow \hat{H}_{0}\left(t\right)-\imath\hat{\Gamma}$ with a corresponding $\hat{H}_F\rightarrow \hat{H}_F-\imath\sum_n
\hat{u}_{n}^{\dagger}\hat{\Gamma}\hat{u}_{n}$.

Finally, the high-frequency expansion of the effective Hamiltonian $\hat{H}_{F}$ is given in many references \cite{LiBT2018,Eckardt2017}, 
and it is shown here for the sake of completeness and convention consistency: 
\begin{widetext}
\begin{align}
\hat{H}_{F} \approx\hat{H}^{\left(0\right)}-\sum_{n=1}^{\infty}\frac{1}{n\omega}\left[\hat{H}^{\left(n\right)},\hat{H}^{\left(-n\right)}\right]+
\sum_{n\neq0}\frac{\left[\hat{H}^{\left(-n\right)},\left[\hat{H}^{\left(0\right)},\hat{H}^{\left(n\right)}\right]\right]}{2n^{2}\omega^{2}}
+\sum_{n\neq0}\sum_{n'\neq0,n}\frac{\left[\hat{H}^{\left(-n'\right)},\left[\hat{H}^{\left(n'-n\right)},\hat{H}^{\left(n\right)}\right]\right]}{3nn'\omega^{2}}.
\label{HF}
\end{align}
\end{widetext}
Hereafter we will consider tight-binding leads with dispersion relation $E\left(k\right)=-2\cos\left(k\right)$ (in units of coupling). In this 
case, $\lambda=\cos\left(k\right)$ (appearing in $\Lambda_u$) and the group velocity $v_{g}=2\sin\left(k\right)$, 
where $k\in\left(0,\pi\right)$.

Finally, we point out the structural similarities between the $S$-matrix given by  Eq.~(\ref{Smat}) and the scattering matrix associated
with static targets, when the latter is written in terms of the Hamiltonian of the corresponding isolated system, see for example 
Refs.~\cite{FS1997,LiST2018}. These similarities can be proven useful in other investigations where scattering properties of complex/
chaotic systems are investigated. Interestingly enough, although the formula Eq.~(\ref{Smat}) is formally correct up to order ${\cal O}
\left(1/\omega^{2}\right)$, its performance typically go beyond this order (compare, for example, Fig.~\ref{fig2}b and Fig.~\ref{fig2}a).
The reason is subtle but we might appreciate it from the preserved structure and general properties. For example, in the absence of 
gain/loss, i.e., $\hat{\Gamma}=0$, the approximate scattering matrix in Eq.~(\ref{Smat}) preserves the unitarity as it should.

Below we will present our design strategy for the realization of non-reciprocal components with {\it minimal complexity}. It consists of 
two steps: first we utilize the Floquet engineering within the scattering set-up to develop driving schemes that provide optimal 
non-reciprocal responses in the high-frequency limit. At the second step we demonstrate that these same designs manifest optimal 
transmittance asymmetry also in the limit of slow driving. Our proposal is supported by detail simulations.

\begin{figure}
\includegraphics[width=1\linewidth, angle=0]{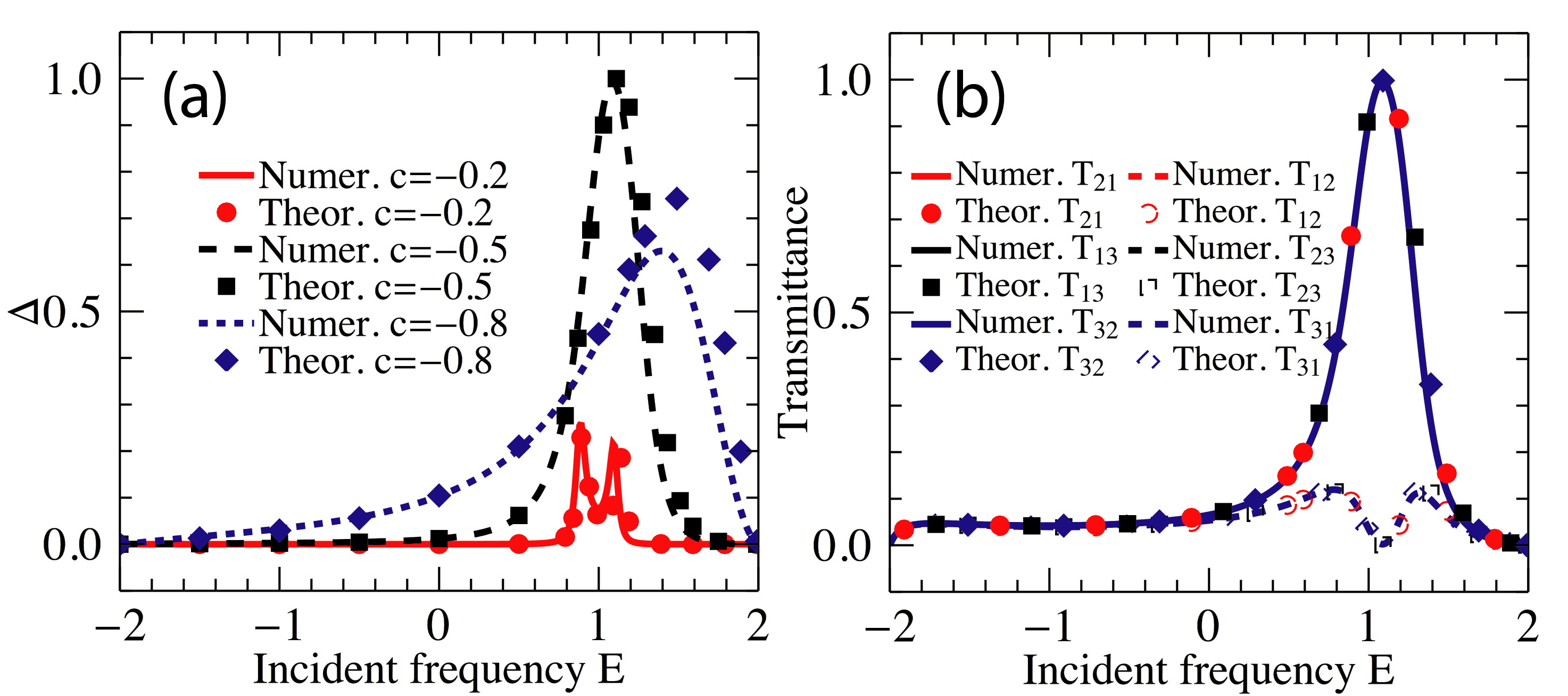}
\caption{(a) Non-reciprocal transmittance $\Delta$ between leads 1 and 2 (see Fig. \ref{fig1}b) for the  model Eq.~(\ref{circulator_example}). 
An optimal phase configuration $\phi_{1}=0$, $\phi_{2}=2\pi/3$ and $\phi_{3}=4\pi/3$ and various coupling constants $c$ are considered. 
Maximum non-reciprocity $\Delta=1$ is achieved for perfect impedance matching conditions corresponding to $c\approx-0.5$ (and $E\approx 
1.1$. The numerical data (lines) have been derived using the (exact) Floquet scattering matrix, see Ref.~\cite{Li2018}. The theoretical results 
(symbols) correspond to Eq.~(\ref{circulator_eg_delta}). (b) The transmittances $T_{nn'}=\left|S_{nn'}\right|^{2}$ in clockwise/counterclockwise 
direction versus the incident frequency $E$ when the coupling $c=-0.5$. The theoretical results (symbols) Eq.~(\ref{Smat}) match perfectly 
the numerical results (lines). The other parameters for both (a) and (b) are: $h_{0}=-1, h=0.5, \varepsilon_0=0$ and $\omega=6$.}
\label{fig3}
\end{figure}


\section{Design of nonreciprocal components with minimal complexity: High frequency limit --} \label{high}

First we utilize Eq. (\ref{Smat}) to derive 
analytical expressions for the left/right transmittance asymmetry $\Delta$ up to leading order in driving frequency $\omega$, i.e. $\Delta
\propto {\cal O}(1/\omega)$. A subsequent optimization of these expressions with respect to various driving parameters, allow us to engineer 
non-reciprocal components (circulators, isolators) with maximum transmittance asymmetry.

\subsection{Reconfigurable Isolators}\label{isolators}

The first reconfigurable nonreciprocal component that we design is an isolator. We consider the simplest possible scenario involving 
two driving modes, i.e., $N_{s}=2$, coupled to two leads $M=2$ ~\cite{note0}. We first observe that the leading order contribution 
$\propto {\cal O}\left(1/\omega\right)$ in 
the expression Eq. (\ref{Smat}) for the scattering matrix, originates solely from the effective Floquet Hamiltonian $\hat{H}_{F}$ (see 
Eq. (\ref{HF})). Due to its Hermitian nature, however, the $2\times2$ matrix $\hat{H}_{F}$ cannot lead to a transmittance asymmetry 
$|S_{12}|\neq |S_{21}|$. One needs, therefore, to ``break" this Hermiticity constraint -- an operation performed by incorporating in 
$\hat{H}_F$ the non-Hermitian diagonal matrix $\imath\hat{\Gamma}$ (see previous discussion). As a result, the matrix $G$ in Eq. 
(\ref{Smat}) has, up to a common factor, complex diagonal matrix elements $G_{11},G_{22}\in {\cal C}$ while its off-diagonal elements 
are complex conjugates of each other $G_{12}=G^*_{21}$. This, by itself, does not guarantee that $|S_{21}|\neq |S_{12}|$, let alone 
that $\Delta\propto {\cal O}\left(1/\omega\right)$. The latter requirement can be fulfilled only when the bare coupling matrix $\hat{c}$ 
has off-diagonal elements which lead to a mixing of diagonal and off diagonal terms of $G$ after performing the multiplication (see 
Eq. (\ref{Smat})). The presence of off-diagonal elements in $\hat{c}$ {\it suggests an isolator design that involves physical loops}. An 
example of such system is shown in Fig. \ref{fig1}a and it is mathematically modeled via the bare coupling matrix $\hat{c}=
\begin{pmatrix}c_{L} & \tilde{c}_{L}\\0 & c_{R}\end{pmatrix}$. We stress again that this design is a direct consequence of the 
theoretical analysis of Eq. (\ref{Smat}). 

Next we proceed with the evaluation of transmittance asymmetry $\Delta$ for the design of Fig.~\ref{fig1}a. For simplicity, we assume 
uniform gain/loss, i.e., $\hat{\Gamma}=\gamma I_{N_s}$. Furthermore, we parametrize the effective Hamiltonian $\hat{H}_{F}$ as 
$\hat{H}_{F} =\hat{H}_{F}^{\dagger}=\begin{pmatrix}\eta_{1} & \eta_{r}+\imath\eta_{i}\\\eta_{r}-\imath\eta_{i} & \eta_{2}\end{pmatrix}$ 
and the micromotion operator $\hat{u}_{0}$ as $\hat{u}_{0}=\hat{u}_{0}^{\dagger}=\begin{pmatrix}\mu_{1} & \mu_{r}+\imath\mu_{i}
\\\mu_{r}-\imath\mu_{i} & \mu_{2}\end{pmatrix}$. In the above parametrization we have consider terms up to order ${\cal O}\left(1/
\omega^{2}\right)$. Specifically, since the target in the absence of modulation is recirpocal, we have that $\eta_{i}\sim {\cal O}\left(
1/\omega\right)$ while generally $\left\{ \eta_{1},\eta_{2},\eta_{r}\right\} \sim {\cal O}\left(1\right)$. Similarly, the components of the 
micromotion operator $\hat{u}_{0}$ are $\mu_{1,2}=1+{\cal O}\left(1/\omega^{2}\right)$ and $\mu_{r,i}\sim {\cal O}\left(1/\omega^{2}\right)$.

It turns out that the matrix elements $\eta_{i}$ and $\mu_{i}$ are essential for the presence of the transmittance asymmetry. Their 
origin are traced back to the presence of driving and can be associated with an effective gauge field \cite{FangFan2012}. To appreciate 
the importance of $\eta_{i}$ and $\mu_{i}$, we have evaluated the transmittance asymmetry $\Delta$ explicitly up to ${\cal O}\left(1
/\omega^{2}\right)$. Using Eq.~(\ref{Smat}) we get 
\begin{align}
\Delta & \approx-\frac{4c_{L}c_{R}^{2}v_{g}^{2}\gamma\left\{ \eta_{i}\tilde{c}_{L}+\mu_{i}\left[\tilde{c}_{L}\left(\eta_{1}-\eta_{2}\right)-2c_{L}
\eta_{r}\right]\right\} }{\left|\det D\left(\gamma\right)\right|^{2}}\nonumber\\
& \propto \eta_i + {\cal O}(1/\omega^2)
\label{isolator_delta}
\end{align}
where the matrix $D\left(\gamma\right)\equiv E+\imath\gamma-\hat{H}_{F}+e^{\imath k}\hat{c}^{\dagger}\hat{c}$ and $\hat{H}_{F}$ 
has been evaluated using the first two terms in Eq. (\ref{HF}). 

Let us now consider a specific driving protocol described by the time-dependent Hamiltonian $\hat{H}_{0}
\left(t\right)$
\begin{align}
\hat{H}_{0}\left(t\right) & =\begin{pmatrix}\varepsilon_{L}\left(t\right) & h\left(t\right)\\
h\left(t\right) & \varepsilon_{R}\left(t\right)
\end{pmatrix}
\label{Eg_isolator}
\end{align}
where $\varepsilon_{L/R}\left(t\right)=\varepsilon_{L/R}^{0}+2f_{L/R}\cos\left(\omega t+\phi_{L/R}\right)$ and $h\left(t\right)=h_{0}+2h\cos
\left(\omega t+\phi_{0}\right)$. Without any loss of generality we will assume that the starting time of the driving scheme is such that $\phi_{0}
=0$. Then the other two driving phases $\phi_{L/R}$ can be measured \textit{relative} to $\phi_0$, and can be used to optimize the transmittance 
asymmetry. For the specific driving protocol of Eq.~(\ref{Eg_isolator}), we have that $\mu_{i}\sim {\cal O}\left(1/\omega^{3}\right)$, $\eta_{i}=
\frac{2h}{\omega}\left(f_{L}\sin\phi_{L}-f_{R}\sin\phi_{R}\right)+{\cal O}\left(1/\omega^{3}\right)$ and $\hat{H}_{F}=\begin{pmatrix}\varepsilon_{L}^{0} 
& h_{0}+\imath\eta_{i}\\h_{0}-\imath\eta_{i} & \varepsilon_{R}^{0}\end{pmatrix}+{\cal O}\left(1/\omega^{2}\right)$. A theoretical expression for the 
transmittance asymmetry $\Delta$, as a function of incident frequency $E$, can be calculated by direct substitution of $\mu_i,\eta_i$ into 
Eq.~(\ref{isolator_delta}). Specifically we find that up to ${\cal O}\left(1/\omega\right)$, the following expression for the transmittance asymmetry 
\begin{equation}
\label{isolator}
\Delta\propto\eta_{i}=\frac{2h}{\omega}\left(f_{L}\sin\phi_{L}-f_{R}\sin\phi_{R}\right)
\end{equation}
which takes its maximum value $\left|\Delta\right|$ for $\phi_{L}=-\phi_R=\pi/2$.

In Fig. \ref{fig2}a we report the theoretical predictions Eq. (\ref{isolator_delta}) for the transmittance asymmetry $\Delta$, together with 
the outcome of the simulations (where the exact Floquet scattering matrix has been used \cite{Li2018}). Various values of the 
driving phases $\phi_L=\pi,\pi/2,\pi/6$ (for fixed $\phi_R=-\pi/2$) have been considered. In all cases we observed a nice agreement between 
theory and numerics. Moreover, we find that the maximal transmittance asymmetry occurs  when $\phi_{L}\approx\pi/2$, as predicted by 
our theory. This optimal phase choice is further verified in the inset of Fig.~\ref{fig1}a where we report $\Delta$ versus $\phi_L$ for a fixed 
incident wave-frequency $E=1$. At Fig. \ref{fig2}b we also report the individual left $T_{L\rightarrow R}$ and right $T_{R\rightarrow L}$ 
transmittances versus the incident wave frequency, for the optimized phase configuration $\phi_{L}=-\phi_{R}=\pi/2$. The theoretical 
results (symbols) have been calculated using the (approximated) expression Eq.~(\ref{Smat}) for the $S$-matrix (high frequency limit).

\subsection{Reconfigurable Circulators}\label{circulators}

We proceed with the design of circulators. As before, we consider a design with the minimal complexity consisting of three (driven) 
modes, each of which is coupled to a lead (i.e., $M=3$) with equal coupling strength $c$. As opposed to the case of (two-channel) isolators, 
here the presence of the non-Hermitian term $\hat{\Gamma}$ (describing material losses/gain) is not necessary. The non-Hermiticity is 
automatically satisfied by the presence of the propagating channel in the third lead and thus we will assume below that $\hat{\Gamma}=0$. 
At the absence of driving, the system is respecting a rotational symmetry. We want to
design a counter-clock-wise circulator, i.e. a three-port structure for which counter-clock-wise transmittances $T_{21},T_{32},T_{13}\neq 
0$, while transmittances in the clock-wise direction are (essentially) zero i.e. $T_{31},T_{23},T_{12}\approx 0$. Obviously, such a structure
must demonstrate a strong {\it positive} transmittance asymmetry $\Delta>0$ between two consequent leads. A schematics of this circulator 
is shown in Fig. \ref{fig1}b.

Similar to the case of the isolator, we parametrize the effective Hamiltonian $\hat{H}_{F}$ as
\begin{align}
\hat{H}_{F} & =\begin{pmatrix}\eta_{01} & \eta_{1r}+\imath\eta_{1i} & \eta_{3r}-\imath\eta_{3i}\\
\eta_{1r}-\imath\eta_{1i} & \eta_{02} & \eta_{2r}+\imath\eta_{2i}\\
\eta_{3r}+\imath\eta_{3i} & \eta_{2r}-\imath\eta_{2i} & \eta_{03}
\end{pmatrix}\label{eq: circulator_P_HF}
\end{align}
where $\eta_{ni}\sim {\cal O}\left(1/\omega\right), \eta_{0,n}=\epsilon_0+{\cal O}(1/\omega), \eta_{n,r}=h_0+{\cal O}(1/\omega)$ and
$n=1,2,3$. At the same time the micromotion operator $\hat{u}_{0}$ can be written as
\begin{align}
\hat{u}_{0} & =\begin{pmatrix}\mu_{01} & \mu_{1r}+\imath\mu_{1i} & \mu_{3r}-\imath\mu_{3i}\\
\mu_{1r}-\imath\mu_{1i} & \mu_{02} & \mu_{2r}+\imath\mu_{2i}\\
\mu_{3r}+\imath\mu_{3i} & \mu_{2r}-\imath\mu_{2i} & \mu_{03}
\end{pmatrix},\label{eq: circulator_P_u0}
\end{align}
with $\mu_{0n}=1+{\cal O}\left(1/\omega^{2}\right)$, $\{\mu_{nr},\mu_{ni}\}\sim {\cal O}\left(1/\omega^{2}\right),n=1,2,3$.

\begin{figure}
\includegraphics[width=1\linewidth, angle=0]{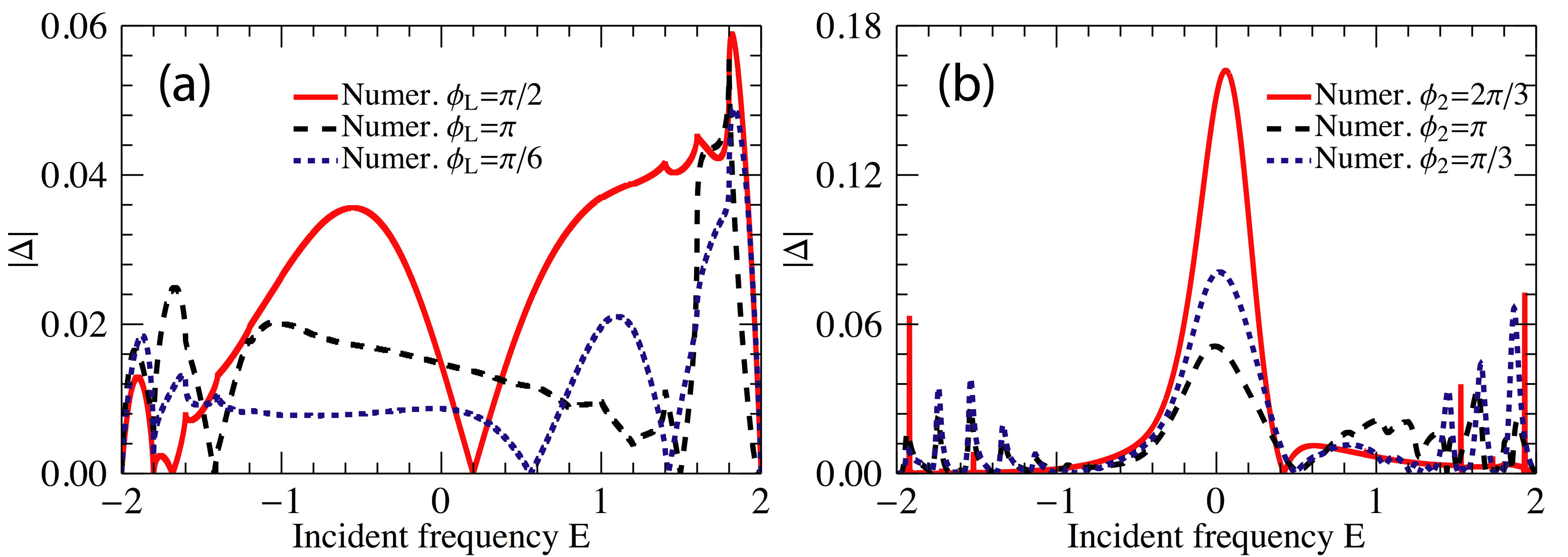}
\caption{Performance of nonreciprocal components in the low driving-frequency limit $\omega=0.2$. (a) Isolator: the magnitude of the 
transmittance asymmetry
$\left|\Delta\right|$ versus the incident frequency $E$ when $\phi_{L}=\pi,$
$\pi/2$ and $\pi/6$ respectively. The model for the isolator and
the other parameters are the same as in Fig.~\ref{fig2}. (b)
Circulator: the magnitude of the transmittance asymmetry $\left|\Delta\right|$
versus the incident frequency $E$ when $\phi_{2}=2\pi/3,$ $\pi$
and $\pi/3$ while $\phi_1=0$ and $\phi_3=4\pi/3$ are kept fixed. The design of the circulator and the other
parameters are the same as in Fig.~\ref{fig3}b. As shown
in the subfigures (a) and (b), the relative driving phases for the (overall)
optimal performance of the nonreciprocal components are consistent
with the high driving-frequency prediction.}
\label{fig4}
\end{figure}

From Eqs.~(\ref{Smat},\ref{eq: circulator_P_HF},\ref{eq: circulator_P_u0}) we are able to derive the following expression for the transmittance 
asymmetry $\Delta$ up to ${\cal O}\left(1/\omega^{2}\right)$ between two consequent leads (say lead 1 and lead 2, see Fig. \ref{fig1}b):
\begin{align}
\Delta & \approx\frac{\left(c^{2}v_{g}\right)^{3}\left(h_{\eta}+m_{\mu}\right)}{\left|\det D\left(0\right)\right|^{2}}\propto h_{\eta}+{\cal O}(1/\omega^2)
\label{circulator_delta}
\end{align}
where the matrix $D(\gamma)$ is defined below Eq.~(\ref{isolator_delta}). Above $h_{\eta}\equiv2\left(\eta_{1r}\eta_{2r}\eta_{3i}+\eta_{3r}
\eta_{1r}\eta_{2i}+\eta_{2r}\eta_{3r}\eta_{1i}\right)$ originates from the effective Hamiltonian $\hat{H}_{F}$ and $m_{\mu}\equiv4h_{0}^{2}
\left(h_{0}-\varepsilon_{0}-2\cos k\right)\sum_{n}\mu_{ni}$ from the micromotion. Due to the structural symmetry of the undriven system, 
Eq.~(\ref{circulator_delta}) applies also for the transmittance asymmetry between the leads $2$ ($3$) and $3$ ($1$).

Let us work out a specific driving protocol described by the following time-periodic Hamiltonian 
\begin{align}
\hat{H}_0\left(t\right) & =\begin{pmatrix}\varepsilon_{0} & h_{1}\left(t\right) & h_{3}\left(t\right)\\
h_{1}\left(t\right) & \varepsilon_{0} & h_{2}\left(t\right)\\
h_{3}\left(t\right) & h_{2}\left(t\right) & \varepsilon_{0}
\end{pmatrix},\label{circulator_example}
\end{align}
where the periodic modulation pertains to the couplings $h_{n}\left(t\right)=h_{0}+2h\cos\left(\omega t+\phi_{n}\right)$. 
We want to identify the driving phases configuration $\{\phi_n\}$ for which the circulator will demonstrate maximum performance.
In this case we have that $\mu_{ni}\sim {\cal O}\left(1/\omega^{3}\right)$, $\eta_{nr}=h_{0}+{\cal O}\left(1/\omega^{2}\right)$, $\eta_{1i}
=\frac{2h^{2}}{\omega}\sin\left(\phi_{3}-\phi_{2}\right)+{\cal O}\left(1/\omega^{3}\right)$ (and cyclicly $\eta_{2i}=\frac{2h^{2}}{\omega}\sin\left(\phi_{1}-\phi_{3}\right)+{\cal O}\left(1/\omega^{3}\right)$, etc) \cite{note2}. According to Eq.~(\ref{circulator_delta}),
the associated transmittance asymmetry reads 
\begin{widetext}
\begin{align}
\Delta & \approx\frac{4h_{0}^{2}h^{2}\left(c^{2}v_{g}\right)^{3}\left(\sin\left(\phi_{2}-\phi_{1}\right)+\sin\left(\phi_{1}-\phi_{3}\right)+
\sin\left(\phi_{3}-\phi_{2}\right)\right)}{\omega\left|\det D\left(0\right)\right|^{2}}.\label{circulator_eg_delta}
\end{align}
\end{widetext}
The above equation indicates that $\Delta$ depends only on the relative driving phases i.e. $\phi_{2}-\phi_{1}$
and $\phi_{3}-\phi_{1}$. Without loss of generality, we set $\phi_{1}=0$. From Eq. (\ref{circulator_eg_delta}) we find that the phase configuration
$\phi_{2}=2\pi/3$ and $\phi_{3}=4\pi/3$ (or similarly $\phi_{2}=4\pi/3$ and $\phi_{3}=2\pi/3$) can produce the maximum asymmetry up to order 
${\cal O}\left(1/\omega\right)$. We point out that a similar driving phase configuration has been implemented recently in the case of mode-modulated
circulators~\cite{Alu2014} - though in our case it is important to realize that this optimal configuration emerged as a result of our optimization
approach. We can further optimize $\Delta$ with respect to the coupling parameter $c$. It turns out that for the specific case of optimal
phases the critical coupling (perfect impedance matching) occurs for $c\approx -0.5$.

In Fig.~\ref{fig3}a, we show the numerical data for the transmission asymmetry $\Delta$ between channels 1 and 2 (see Fig. \ref{fig1}b) 
versus the incident frequency for the optimal phase configuration $\phi_{1}=0$, $\phi_{2}=2\pi/3$ and $\phi_{3}=4\pi/3$. Various coupling 
strengths $c$ have been considered. At the same figure we also report the theoretical results for $\Delta$, see Eq.~(\ref{circulator_eg_delta}). 
We found a non-monotonic behavior of the transmittance asymmetry with respect to the coupling $c$ due to the impedance mismatch. 
For the predicted coupling $c\approx-0.5$, the system demonstrates a nonreciprocal behavior which is as high as $100\%$. 

The individual transmittances $T_{nn'}=\left|S_{nn'}\right|^{2}$ versus incident frequency $E$ (for the optimal phase and coupling 
configurations) are reported in Fig.~\ref{fig3}b. Both theoretical (using Eq. (\ref{Smat})) and numerical (using the exact Floquet scattering 
matrix \cite{Li2018}) results fall nicely one on top of the other and indicate that for the optimal phase configuration $T_{31}=T_{23}=
T_{12}\approx 0$ while $T_{21}=T_{32}=T_{13}\neq 0$ at a frequency range around $E\approx 1.1$.

\section{Nonreciprocal components with minimal complexity: Low driving frequency limit}\label{low} 

Although the Floquet $S$-matrix can be always 
expressed in terms of ${\hat H}_F$, one cannot use any more the approximated forms Eqs.~(\ref{Smat},\ref{u0},\ref{HF}) in the slow-frequency 
driving limit. As a result, our theoretical expressions Eqs.~(\ref{isolator_delta},\ref{circulator_delta}) for the transmittance asymmetry are not
any more applicable. We have found, nevertheless, that the driving designs that lead to optimal non-reciprocity in the case of high-frequency 
driving schemes are applicable even in the case of small driving frequencies. This conclusion has been supported via detail numerical 
simulations for various drivings. Typical examples are shown in Fig. \ref{fig4}a,b where we plot the numerical results for $|\Delta|$ (using 
the exact Floquet scattering matrix \cite{Li2018}) for the previous designs of isolators and circulators. A small driving frequency 
$\omega=0.2$ is now used. In the case of an isolator, see Fig. \ref{fig4}a, we report $|\Delta (E)|$ for three representative driving phases $\phi_L=\pi, 
\pi/2, \pi/6$. An overall optimal asymmetric transmittance is observed for $\phi_L=\pi/2$ which is the predicted optimal phase in the case of 
high-frequency driving schemes (see Fig. \ref{fig2}). Similarly, in Fig. \ref{fig4}b we report the behavior of $|\Delta(E)|$ for $\phi_1=0,\phi_3=4\pi/3$
and various values of $\phi_2=2\pi/3,$ $\pi$ and $\pi/3$. Again we find that an overall optimal transmittance asymmetry is observed when
$\phi_2=2\pi/3$ which is the predicted optimal driving phase in the case of high-frequency driving schemes. 

\section{Conclusion}\label{conclusions}

We have developed a scheme for designing optimal non-reciprocal components (isolators, circulators etc) which utilize
the concept of Floquet engineering. In the high-frequency modulation limit we developed a theoretical formalism that relies on the notion 
of effective Floquet Hamiltonians ${\hat H}_F$ and allows us to express the transmittance asymmetry in inverse powers of the modulation 
frequency. An analysis of the theoretical expressions provides guidance for the geometrical design of non-reciprocal components 
and driving schemes that can lead to optimal performance. Specifically, the method highlights the importance of physical loops in 
getting maximum non-reciprocal efficiency. Detail numerical investigation indicates that these designs can provide optimal non-reciprocal 
transport even in the case of low driving frequencies.   

Our approach can find promising applications in the framework of classical electromagnetic and acoustic wave theories as well as 
matter wave physics. Another potential application could be in the framework of thermal transport where the design of reconfigurable 
thermal diodes is a challenging research direction.

\section{Acknowledgments} 

The authors acknowledge useful discussings with Prof. B. Shapiro. This research was partially supported by DARPA NLM program 
via grant No. HR00111820042, by an AFOSR grant No. FA 9550-10-1-0433, and by an NSF grant EFMA-1641109. The views and 
opinions expressed in this paper are those of the authors and do not reflect the official policy or position of the U.S. Government.



\begin{thebibliography}{10}
\bibitem{note0}We remark that one-mode driving can not lead to isolation with $O\left(1/\omega\right)$ performance, which is obvious when
setting $M=2$ in Eq.~(\ref{Smat}) and considering that both the effective Hamiltonian $\hat{H}_{F}$ and the micromotion $\hat{u}_{n}$
are scalar in this case. Notice that we exclude the possibility of the lead-lead coupling, which is not captured by Eq.~(\ref{Smat})\cite{LiBW2012}.
\bibitem{note2}For the example Eq.~(\ref{Eg_isolator}) (and
similarly the previous one (\ref{circulator_example})), the effect
of the micromotion is absent within order $O\left(1/\omega^{2}\right)$.
It will be interesting to work in the the complex hermitian Hamiltonian
$\hat{H}_{0}\left(t\right)$ to investigate the interplay between
the effective Hamiltonian $\hat{H}_{F}$ and the micromotion within
order $O\left(1/\omega^{2}\right)$, which, however, is beyond the
current purpose to design $O\left(1/\omega\right)$ asymmetry in the
nonreciprocal components.
\end{thebibliography}
\end{document}